\documentstyle[12pt,a4,psfig]{article}
\newcommand{\npar}{\bigskip\par\noindent}
\newcommand{\mnu}{m_{\nu}}

\newcommand{\mw}{m_W}
\newcommand{\vsig}{\overline{v_{rel}\,\sigma}}
\begin{document}

\vspace*{1.0cm}
\leftline{\large{DESY 97--057}}
\leftline{\large{April 1997}}

\vspace*{1.5cm}

\bigskip
\centerline{\Large{Cosmic Ray Constraints On the Annihilation}}
\centerline{\Large{of Heavy Stable Neutrinos in the Galactic Halo}}

\vspace*{1.0cm}
\centerline{\large{Yu.A.Golubkov$^{a)}$, R.V.Konoplich$^{b)}$}}

\vspace*{1.0cm}
$^{a)}$ {\em Moscow State University, Moscow, Russia}
{\footnote{e-mail:
golubkov$@$elma01.npi.msu.su,~golubkov$@$vxdesy.desy.de}}.

$^{b)}$ {\em Moscow Engineering Physics Institute, Moscow, Russia}
{\footnote{e-mail: konoplic$@$rosti.mephi.msk.su}}.

\vspace*{1.0cm}
\begin{abstract}
\bigskip
\npar We carry out a detailed analysis of fluxes of cosmic ray antiprotons,
positrons, electrons and photons to be expected from the annihilation 
of relic heavy neutrinos in the galactic halo. The spectra of particles 
are evaluated by Monte Carlo simulation. The results of calculations 
show that the investigation of cosmic ray positron spectra at high 
energies could provide a distinctive signal for annihilation of very 
heavy neutrinos in the Galaxy and give an important information on 
parameters of dark matter particles.
\end{abstract}

\bigskip
\section{Introduction}

\npar
The agreement that more or less exists between the observed and predicted
abundances of light elements such as $D$, $^3He$, $^4He$ and $^7Li$ provides
a confirmation of the standard cosmology and leads to the determination of
the baryon density in the range \cite{Copi95}

\begin{equation}
\label{bardens}
\rho_B\ =\ (1.7\ -\ 4.1)\cdot 10^{-31}\ g/cm^3.
\end{equation}

\npar
Because the critical density

\begin{equation}
\label{critdens}
\rho_c\ =\ 3H_0^2/8\pi G\ =\ 1.88\,h^2\cdot 10^{-29}\ g/cm^3
\end{equation}

\npar
depends on the Hubble constant $H_0$ ($h$ is the Hubble constant in units
of {\hbox{$100\ km\,s^{-1}\,Mpc^{-1}$}}) the present ratio of the baryon
density of the Universe to its critical density also depends on $H_0$,

\begin{equation}
\label{presdens}
\Omega_B\ =\ \rho_b /\rho_c\ =\ (2.2\ -\ 9.0)\,10^{-2}\,h^{-2}.
\end{equation}

\npar
Therefore for a wide range of the Hubble constant values, $h\,\approx\,(0.4\,-\,1)$,
baryons contribute $(1\,-\,15)$\% of the critical density of the Universe. 
Since the contribution of luminous matter

\begin{equation}
\label{lumcontr}
\Omega_{lm}\ \approx\ 3\cdot10^{-3}\,h^{-1}
\end{equation}

\noindent is much less than 1\% of the closure density, most baryons 
in the Universe
should be dark either in the form of hot gas in clusters of galaxies or
in the form of dark stars.

\npar
There are increasing observational evidences on a variety of scales favouring
an average density of matter in the Universe which is significantly
greater than 15\% of the critical density possibly contributed by baryons.
The non--Keplerian character of rotation curves around spiral galaxies
indicates that they are at least ten times more massive than 
the sum of the stars which are being seen \cite{Fich91}. 
The velocity dispersion of stars and $X$--ray emission
in elliptical galaxies show that these objects are also dominated by a dark
component. At the scale of clusters of galaxies, independent sets of
measurements point at an even greater amount of the dark matter. If this is
the case, most of the matter in the Universe must be nonbaryonic.
The evidence for dark matter in galaxies is collected in
Ref.\cite{Ashman92},
and the evidence for dark matter in the Universe is summarized
in Ref.\cite{Trimble89}.

\npar
Nonbaryonic dark matter can be divided into two broad classes on the basis
of its clustering properties --- the hot dark matter being relativistic
at galaxy formation time and the cold one being slowly moving at that epoch.
The most favoured hot dark matter candidate is a light neutrino with a mass 
$\sim\,10\,eV$.
Since light neutrinos are relativistic until relatively recent epochs,
they would free--stream out off and damp out density pertubations up to the
scale of galaxy clusters. However, the results from {\em COBE} 
show that the observed large scale structure cannot be explained 
with purely hot dark matter \cite{Blitz95}.

\npar
The cold dark matter on the another hand clusters at all scales,
since it must have negligible free--streaming length.
In the cold dark matter models,
the structure generally forms hierarchically, with smaller clumps merging
to form the larger ones. For this reason, theorists used to prefer the
cold dark matter scenario over the hot one. But it is worth to note the
recently surging 
popularity of the mixed scenario: a combination of 70\% cold dark matter and
30\% hot one produces a favourable spectrum of density pertubations
and leads to consistency between the inflationary predictions and some
dynamical estimations \cite{Holtzman89}. 
Theoretically, the favourite candidates for cold dark matter
are the axion, an ultra--light pseudoscalar particle, and weakly interacting
massive particles (WIMPS), with masses generally of the order of $100\
GeV$ \cite{{Turner90}, {Raffelt90}}.
The most attractive WIMPS candidate is the neutralino --- 
a supesymmetric fermionic partner of the Standard Model bosons.
For a review of the supersymmetric dark matter see \cite{Jungman96}.
In some models heavy neutrinos play also an important role in cosmology
contributing part of the closure density \cite{{Vysot70},{Lee77},{Hut77}}.

\npar
The nature of the dark matter is certainly one of the most 
important questions facing both particle physics and cosmology.
An indirect way to detect such dark matter particles is to study their
signatures in cosmic rays.

\npar
In the present paper we carry out a detailed analysis of the influence
of the annihilation of very heavy neutrinos ($\mnu\,>\,45\ GeV$)
on cosmic ray production in the Galaxy. There exist a number of scenarios
\cite{{Hill90},{Datta94}} in which such very heavy neutrinos are expected 
to occur. To be explicit, we consider the standard electroweak model, 
including, however, one additiona fermion family.
The heavy neutrino $\nu$ and the heavy charged lepton $L$ form a standard
$SU(2)_L$ doublet. In order to ensure the stability of the heavy 
neutrino $\nu$ 
we assume that $\mnu\,<\,m_L$ and that the heavy neutrino is a Dirac neutrino.

\npar
The organization of the paper is the following. In Sec.2 we discuss 
the residual
relic number density of heavy neutrinos in the Universe and in the Galaxy.
In Sec.3 we present various detailes of our calculations.
In Sec.4 we discuss the results for the spectra of electrons, positrons,
antiprotons and photons resulting from neutrino annihilation 
in the galactic halo
and discuss signatures for the detection of heavy neutrino
in cosmic rays. In Sec.5 we summarize the results of our analysis
of cosmic ray spectra from neutrino annihilation.
In the Appendix we present formulae for matrix elements of the reactions
considered in the paper.

\section{Relic number density of heavy neutrinos}
\label{concent}

\npar
It is known that the available experimental results do not preclude the
existence of heavy Dirac neutrinos with masses $\mnu\,>\,m_Z/2$
\cite{Rev96}. In the early Universe at high temperatures ($T\,>>\,\mnu$)
such heavy neutrinos (if they exist) should be in thermal equilibrium
with other species of particles. As the temperature in the Universe drops,
heavy neutrinos become nonrelativistic at $T\,\sim\,\mnu$ and their
equilibrium concentration is proportional to
$n_{\nu}\,\sim\,exp(-\mnu/T)$ \cite{Dolg81}. 
However, in the further expansion
of the Universe, as the temperature drops below the freeze--out value $T_f$,
the weak interaction processes become too slow to keep neutrinos in
equilibrium with other particles.
As a consequence, the number density
of heavy neutrinos fails to follow the equilibrium concentration 
(the exponential drop of the number density becomes much slower) 
reaching at present \cite{Fargion95}:

\begin{equation}
\label{nuconc}
n_{\nu}(T)\ \approx\ \frac{2\cdot10^{-18}}{\sqrt{g_{\ast}}}\,
 \frac{M_p}{\mnu}\,\frac{1}{\vsig\,M_p^2}\,
\,\left [
40\,+\,\ln \left (
\frac{g_s}{\sqrt{g_{\ast}}}\frac{\mnu}{M_p}
\,\vsig\,M_p^2
\right )
\right ]\,n_{\gamma}(T),
\end{equation}

\noindent where $n_{\gamma}(T)\,=\,0.24T^3$ is the equilibrium 
photon number density, $T\,=\,2.7\ K$ is the present photon temperature,
$M_p$ is the proton mass, $v_{rel}$ is the relative $\nu\bar{\nu}$
velocity in units of the velocity of light. The bar over quantity
$v_{rel}\sigma$ means the averaging over the velocity distribution 
of the heavy neutrino.

\npar
The freeze--out temperature $T_f$ can be found from the 
equation \cite{Dolg81}:

\begin{equation}
\label{tfreez}
\ln\left [ \sqrt{\frac{\mnu}{T_f}}\,
\exp\left (\frac{\mnu}{T_f} \right ) \right ]\ \approx\ 
40\,+\,\ln \left (
\frac{g_s}{\sqrt{g_{\ast}}}\frac{\mnu}{M_p}
\,\vsig M_p^2
\right ),
\end{equation}

\npar
In the above expressions $g_s$ is the num\-ber of par\-ticle 
spin states (for pho\-tons 
and mas\-sive fermi\-ons $g_s\,=\,2$).
$g_{\ast}(T)\,=\,N_{bos}\,+\,\frac{7}{8}N_{ferm}$
is the number of effective degrees of freedom at given temperature
(see \cite{Dolg81}), $N_{bos}$ is the number of bosons and $N_{ferm}$
is the number of fermions (without heavy neutrinos) being in
the thermal equilibrium at given temperature.

\npar 
It was assumed in Eq.(\ref{nuconc}) that 
neutrinos are nonrelativistic at freeze--out, $T_f\,<<\,\mnu$.
The numerical solution for Eq.(\ref{tfreez}) gives $T_f\,=\,3.90\ GeV$
and $T_f\,=\,10.7\ GeV$ for the heavy neutrino masses equal to
$\mnu\,=\,100\ GeV$ and $\mnu\,=\,300\ GeV$, respectively.
Thus, the non--relativistic approximation is valid.
Therefore we can neglect the dependence of the annihilation
cross section on velocity of the neutrino and in further calculations
we take $\sigma(\sqrt{s})\,=\,\sigma(2\,\mnu)$. In this case we can replace
$\vsig\,\approx\,\overline{v_{rel}}\,\sigma(2\,\mnu)$. 
Because $\overline{v_{rel}^2}\,=\,2\,\overline{v^2}$ for definitness
in further calculations we can put
$\overline{v_{rel}}\,=\,\sqrt{2}\,\overline{v}$, where
$\overline{v}\,\approx\,300\ km\cdot s^{-1}$ is the average
velocity of heavy neutrinos in Galactic halo.

\npar
We have to note that Eq.(\ref{nuconc}) has been obtained using
the analytical solution \cite{Dolg81} for the expansion equation 
of hot equilibrium plasma. Such an approach
gives correct order of magnitude, but, in principle, the equation
has to be solved numerically to obtain more precise results. 
In the present paper we do not do this because our main aim is 
to show influence of the hadronization of final state quarks on the 
$e^{\pm},\,p\,\bar p$ and $\gamma$ fluxes as well as to correctly
take into account energy losses for charged final state particles
travelled through the Galaxy.

\npar
In the general case the following processes could lead to the 
annihilation of heavy neutrinos in the early Universe:

$$
\nu\,\bar{\nu}\ \rightarrow \left \{
\begin{array}{lll}
 f\,\bar f & - & \mbox{$s$ channel $Z^0$, $H^0$ exchange}\\
 W^+\,W^-  & - & \mbox{$s$ channel $Z^0$, $H^0$ and
$t$ channel heavy lepton $L$ exchanges}\\
 H^0\,H^0  & - & \mbox{$s$ channel $H^0$ and $t$ channel 
heavy neutrino $\nu$ exchanges}\\
 Z^0\,Z^0  & - & \mbox{$s$ channel $H^0$ and $t$ channel 
heavy neutrino $\nu$ exchanges}\\
 Z^0\,H^0  & - & \mbox{$s$ channel $Z^0$ and $t$ channel 
heavy neutrino $\nu$ exchanges}\\
\end{array}
\right .
$$



\npar
The annihilation of heavy neutrinos into a final state containing $Z^0$ boson
or/and $H^0$ meson are suppressed in comparison with $W^+W^-$ final
state and will be neglected in the following consideration.

\npar
Qualitatively, the dependence of the present neutrino number density
on its mass $\mnu$
can be easily understood if one takes into account that neutrinos at 
freeze--out are nonrelativistic and $\sigma$ depends only on
the mass of the heavy neutrino. The rate of ``burning--out''
of heavy neutrinos is proportional to the rate of collisions
$N\,\sim\,v\,\sigma\,n_{\nu}^2$.
Using Eq.(\ref{nuconc}) we obtain
\(N\,\sim\,1/\sigma\) at $T\,\sim\,T_f$.

\npar
Thus the concentration of the heavy neutrinos at present,
roughly speaking, depends inversely on its annihilation cross section.
Therefore, in the region $\mnu\,\sim\,M_Z/2$, the neutrino
number density is extremely small as a result of the huge value of the
annihilation cross section at the $Z^0$ boson pole. When the
neutrino mass increases the cross section for neutrino annihilation 
into fermions
drops and this leads to an increase of the neutrino number density,
which reaches about $3\cdot10^{-10}\ cm^{-3}$ at $\mnu\,\sim\,100\ GeV$.
At $\mnu\,>\,\mw$ the annihilation channel into $W^+W^-$ opens 
and gradually becomes the dominant one since its cross section 
grows like $\mnu ^2$ \cite{Enqu89} and the present $\nu$ number 
density drops again. 
As it was shown in \cite{Fargion95}
the neutrino number density at $\mnu\,>\,2\,\mw$ is given,
approximately, by

\begin{equation}
\label{apprnnu}
n_{\nu}\ \approx\ \frac{2\cdot10^{-9}}
{\left (\mnu/\mw\right )\,
\left [ \left (\mnu/\mw\right )^2+4\right ]}\,cm^{-3}.
\end{equation}

\npar
Above we discussed the neutrino number density in the Universe. 
However, at the stage of the formation of the Galaxy the motion of heavy
neutrinos in the nonstatic gravitational field of ordinary matter,
which contracts as a results of energy dissipation via radiation,
provides an effective mechanism of energy dissipation for neutrinos
\cite{Zeld80}. As a consequence, the contracting ordinary matter
induces the collapse of the neutrino gas and leads to the following
significant increase in the neutrino density in the Galaxy
\cite{{Fargion95},{Zeld80}}:

\begin{equation}
\label{enhdens}
n_{\nu G}\ \approx\ n_{\nu}\,\left (\frac{\rho_G}{\rho}\right ),
\end{equation}

\noindent where, $\rho_G\,\approx\,5\cdot10^{-24}\ g/cm^{-3}$ is the
average density of matter in the Galaxy and
$\rho\,\approx\,4\cdot10^{-31}\ g/cm^{-3}$
is the density of matter in the Universe.

\section{Fluxes of particles from the heavy neutrino annihilation in the halo}
\label{fluxes}

\npar
The annihilation of dark matter particles in the galactic halo could produce
detectable fluxes of elementary particles such as electrons, positrons,
antiprotons and photons. For the case of supersymmetric dark matter particles 
a number of calculations of cosmic ray fluxes have been made (see, e.g.,
\cite{{Ellis88},{Jung94},{Diehl95}}). In this Section we present
the results of our numerical calculations for fluxes of particles
which appear either directly in the heavy neutrino annihilation or
as secondary products of the fragmentation of quarks and gluons and 
from the decays of unstable particles.

\npar
Let us consider first the fluxes of electrons and positrons from heavy neutrino
annihilation. In this case the differential spectrum of electrons (positrons) 
observable at the Earth is defined by \cite{Zeld80}

\begin{equation}
\label{elflux}
J_e\ =\ \frac{\vsig}{4\,\pi}\,n_{\nu}^2\,\frac{dn_e}{dE}\,v_e\,\tau_e,
\end{equation}

\noindent where $\tau_e$ is the confinement time for electrons (positrons)
in the galactic halo equal to \hbox{$\tau_e\,\approx\,10^7$ years} and
$v$ is the interstellar $e^-(e^+)$ velocity. 
The quantity $dn_e/dE$ is the energy spectrum of electrons produced 
in the annihilation of the heavy neutrino and antineutrino and normalized 
on one collision;
$n_{\nu}$ is the present density of relic heavy neutrinos in the
galactic halo.

\npar
In our calculations we assume that \hbox{$\tau_e\,\approx\,10^7$ years}, but
we have to note that the uncertainties here are substantial and can reach
an order of magnitude. Moreover, $\tau_e$ can, in principle, 
depend on the initial energy of electron.
Additional complications arise since the electrons (positrons) trapped
in the galactic halo lose energy due to various processes.
The energy losses for electrons and positrons propagating 
through the Galaxy halo
matter are being defined by \cite{{Ellis88},{Longair81}}:

\begin{equation}
\label{losses}
\frac{dE}{dt}\ =\ -\left [ 2.7\,n_H\,+\,7.3\,n_H\,E\,
+\,1.02\,\left (U_{rad}\,+\,U_{mag}\right )\,E^2\right ]\,\times\,
10^{-16}\ GeV\,s^{-1},
\end{equation}

\noindent where $U_{rad}$ is the ambient photon energy density 
in $eV\,cm^{-3}$ and $E$ is the $e^-(e^+)$ energy in $GeV$. 
The four terms in (\ref{losses}) represent 
the energy losses due to ionization of the $H$ atoms, 
bremsstrahlung, inverse Compton scattering
and synchrotron radiation in the galactic magnetic field, respectively. 
The electrons (positrons) spend most of their time,
and lose most of their energy in the halo outside of the disc, where
\hbox{$n_H\,\sim\,0.01\ atoms\ cm^{-3}$}, 
\hbox{$U_{rad}\,\sim\,0.25\ eV\ cm^{-3}$} and
\hbox{$U_{mag}\,\sim\,0.05\ eV\ cm^{-3}$}, so that inverse Compton scattering is
dominant for $E\,\ge\,1\ GeV$. The $e^-$ and $e^+$ spectra produced in
neutrino annihilation must be corrected for these energy losses
before the comparison with observations.

\npar
To calculate the particle fluxes from
annihilation of very heavy neutrinos in the Galaxy we used the Monte Carlo
approach. Such an approach allows to take into account the production
of the particles of interest ($e^+,\,e^-,\,\gamma,\,p,\,\bar p$) 
by hadronization of initial quarks or decays of heavier states.
In this approach it is also easy to take into account the energy losses of
final state electrons and positrons in the Galaxy.

\npar
Because we consider large masses of the heavy neutrino the resulting
energies
of final particles are also large, $E\,>\,1\ GeV$. In this case the
dominant contribution comes from the two last terms 
in Eq.(\ref{losses}), which are proportional to $E^2$,
so that, $dE\,=\,-a\,E^2\,dt$, 
where $a\,=\,1.02\cdot\,10^{-16}\,(U_{rad}+U_{mag})$.
Therefore, to simulate the broadening effect of energy losses we 
can use the distribution followed from the simple, 
spatially homogeneous model \cite{{Longair81},{Wilczek90}}:

\begin{equation}
\label{mclosses}
F(E,t)\ \sim\ \exp\left (-\,\frac{1}{a\,t\,E}\right ),
\end{equation}

\noindent which leads to:

\begin{equation}
\label{simlosses}
E\ =\ \frac{E_0}{1\,-\,1.02\,\left (U_{rad}\,+\,U_{mag}\right )
\,10^{-16}\,E_0\,\tau_e\,\ln\xi},
\end{equation}

\noindent where $E$ is the final energy of particle, $E_0$ is its initial
energy and $\xi$ is the random variable, with uniform distribution in the 
$(0,1)$.

\npar
As it was mentioned in Sec.\ref{concent} we restricted our consideration 
to the following two processes:

\begin{equation}
\label{react}
\begin{array}{lll}
\nu\ +\ \bar \nu & \rightarrow & f\ +\ \bar f, \\
\nu\ +\ \bar \nu & \rightarrow & W^+\ +\ W^-, \\
\end{array}
\end{equation}

\noindent where $f$ denotes a lepton or a quark. The expressions
for the differential cross section and matrix elements are given
explicitly in the Appendix.

\npar
The simulation of the annihilation processes 
were perfomed with the package {\em PYTHIA 5.7} \cite{Pythia57} with
suitable modifications to include the fourth generation of fermions. 
To perform the hadronization and decays of final states we used the package 
{\em JETSET 7.4} \cite{Pythia57}. 
This package is well tuned to the experimental data and gives
a satisfactory description of final state multiplicities and particle
distributions.
We assumed that there is no mixing between the three light
generations and the fourth one.
For the hadroniztion of light quarks and gluons we used the symmetric LUND
fragmentation functions and for heavy quarks the Peterson function.
We also took into account the final state radiation in the annihilation 
reaction. This leads to a drastical increase of the spectra in the
very low energy region.
We have taken into account the decays of $\pi$, $n$ and $\mu$, which 
particles are usually considered to be stable.
To avoid unnecessary complications we chose very high masses
for the heavy lepton and Higgs $m_L\,=\,1000\ GeV$, $m_H\,=\,10000\ GeV$,
because, as we have already mentioned previously, the Higgs contribution is
important only in the resonance region $\mnu\,\approx\,m_H/2$.
We performed the simulation for two mass values of the heavy neutrino:
$\mnu\,=\,100\ GeV$ and $\mnu\,=\,300\ GeV$
assuming for the mass of the $t$ quark $m_t\,=\,170\ GeV$. The other 
parameters have been left equal to their default values used in the 
{\em PYTHIA 5.7/JETSET 7.4} package (see for detailes \cite{Pythia57}).

\npar
The results of our calculations for particle spectra are presented 
in Figs.\ref{mnu100}, 
\ref{mnu300} and for total cross sections in Table.1.

\section{Discussion of numerical results}
\label{resdisc}

\begin{figure}[htb]                
\par
\centerline{\hbox{%
\psfig{figure=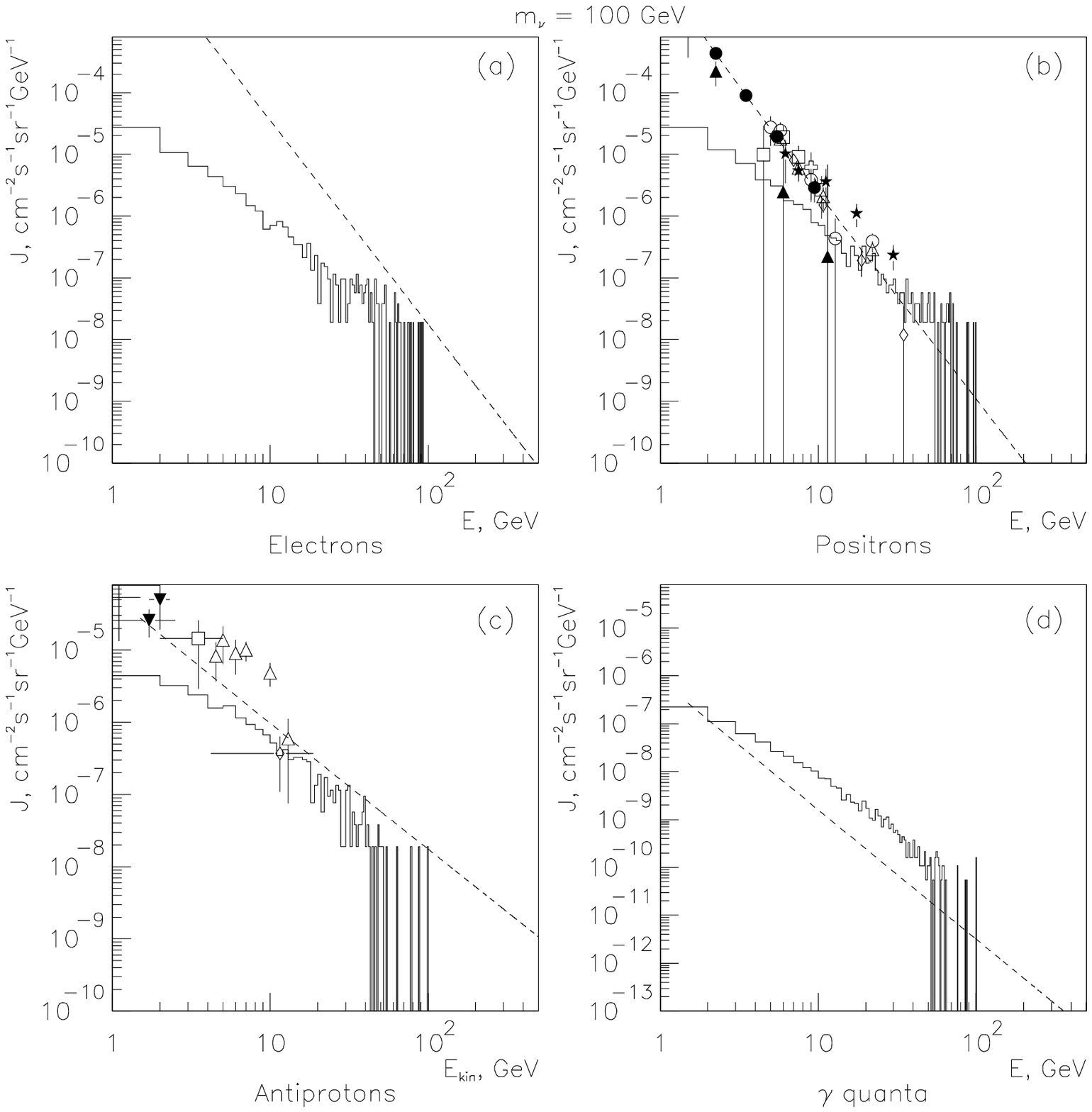,bbllx=1.5cm,bblly=5.5cm,%
bburx=19.0cm,bbury=23.0cm,clip=t,height=16.0cm}%
}}
\par
\caption{\label{mnu100}
Results of the simulation for $\mnu\,=100\,GeV$.
(a) --- electrons; (b) --- positrons; (c) --- antiprotons 
and (d) --- $\gamma$'s.
The histograms present the predicted particle fluxes for 
an energy resolution 
$\Delta E\,=\,1\ GeV$. 
The dashed lines show the observable fluxes 
for present day experiments (see text).
}
\par
\end{figure}

\begin{figure}[htb]                
\par
\centerline{\hbox{%
\psfig{figure=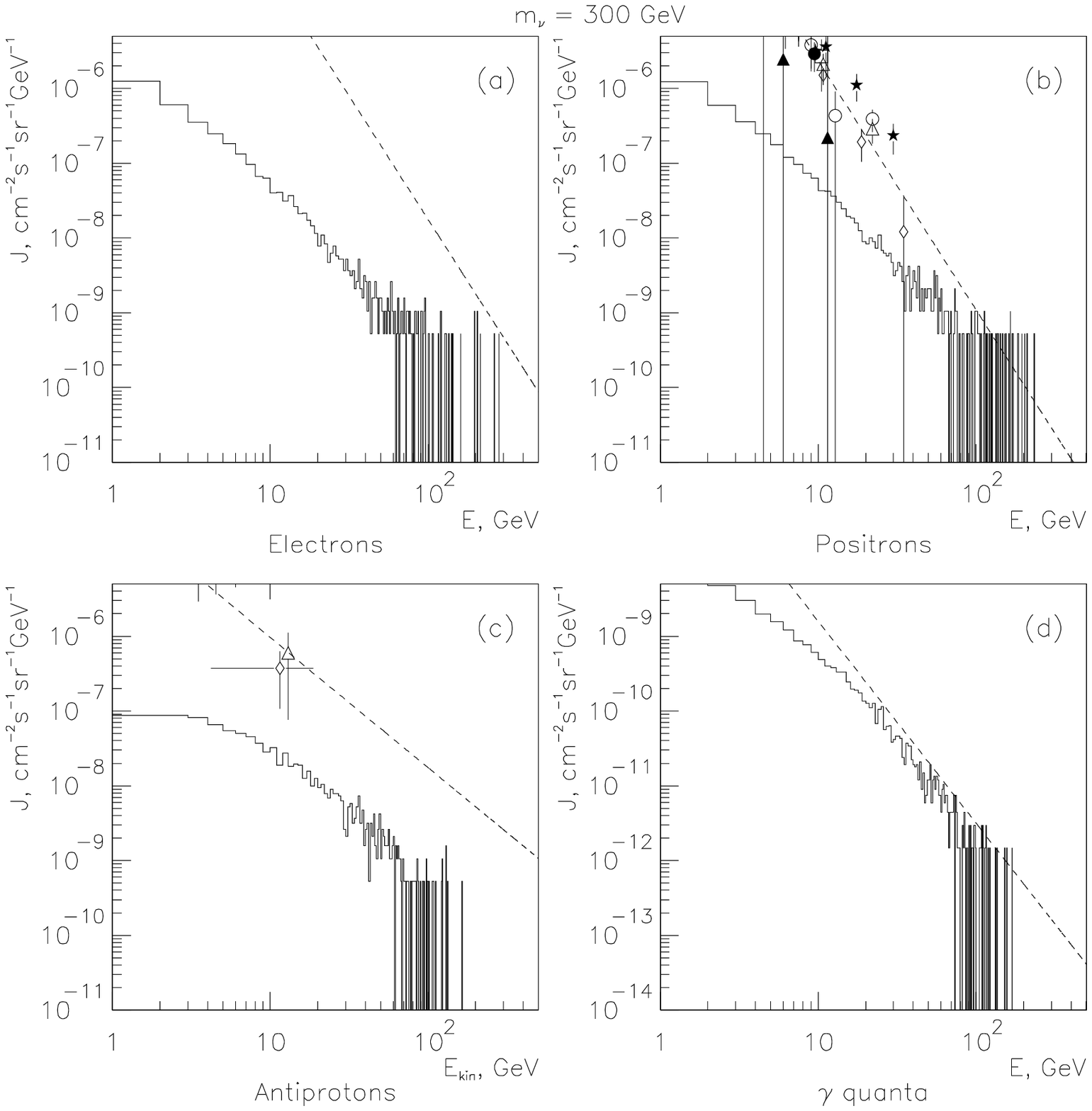,bbllx=1.5cm,bblly=5.5cm,%
bburx=19.0cm,bbury=23.0cm,clip=t,height=16.0cm}%
}}
\par
\caption{\label{mnu300}
Results of the simulation for $\mnu\,=300\,GeV$.
(a) --- electrons; (b) --- positrons; (c) --- antiprotons 
and (d) --- $\gamma$'s.
The histograms present the predicted particle fluxes for
an energy resolution $\Delta E\,=\,1\ GeV$. 
The dashed lines show the observable fluxes 
for present day experiments (see text).
}
\par
\end{figure}

\npar
The results of the numerical simulation of the electron flux are presented
in Fig.\ref{mnu100}(a) and in Fig.\ref{mnu300}(a) for $\mnu\,=\,100\ GeV$ 
and $\mnu\,=\,300\ GeV$, respectively, together with an approximation
of the observable electron spectrum \cite{Mueller87} of the type:

\begin{equation}
\label{expelec}
J_e^{exp}\ =\ 0.07\,E^{-3.3}\ \ \ cm^{-2}\,sr^{-1}\,s^{-1}\,GeV^{-1},
\end{equation}

\noindent where $E$ is $e^-(e^+)$ energy in $GeV$.

\npar
Without energy losses \cite{Fargion95} we could expect a very narrow
peak at $E_e\,=\,\mnu/2$ with width $\Delta E_e\,\sim\,m\,v/c$ well above
the background from the annihilation process
$\nu\,\bar{\nu}\,\rightarrow\,e^+e^-$
(since heavy neutrinos in the Galaxy are nonrelativistic,
$v\,\approx\,300\ km/s$
is the velocity of neutrinos in the Galaxy) and from the continuum 
electron radiation resulting
from decays of unstable particles. However, as we can see from 
Fig.\ref{mnu100}(a) and Fig.\ref{mnu300}(a), due to energy losses 
the resulting electron spectrum becomes significantly softer with a
sharp edge at $E\,=\,\mnu$.
It appears that the electron spectrum
cannot provide a clear signature for the heavy neutrino annihilation
in the galactic halo.

\begin{figure}[htb] 
\centerline{\hbox{
\psfig{figure=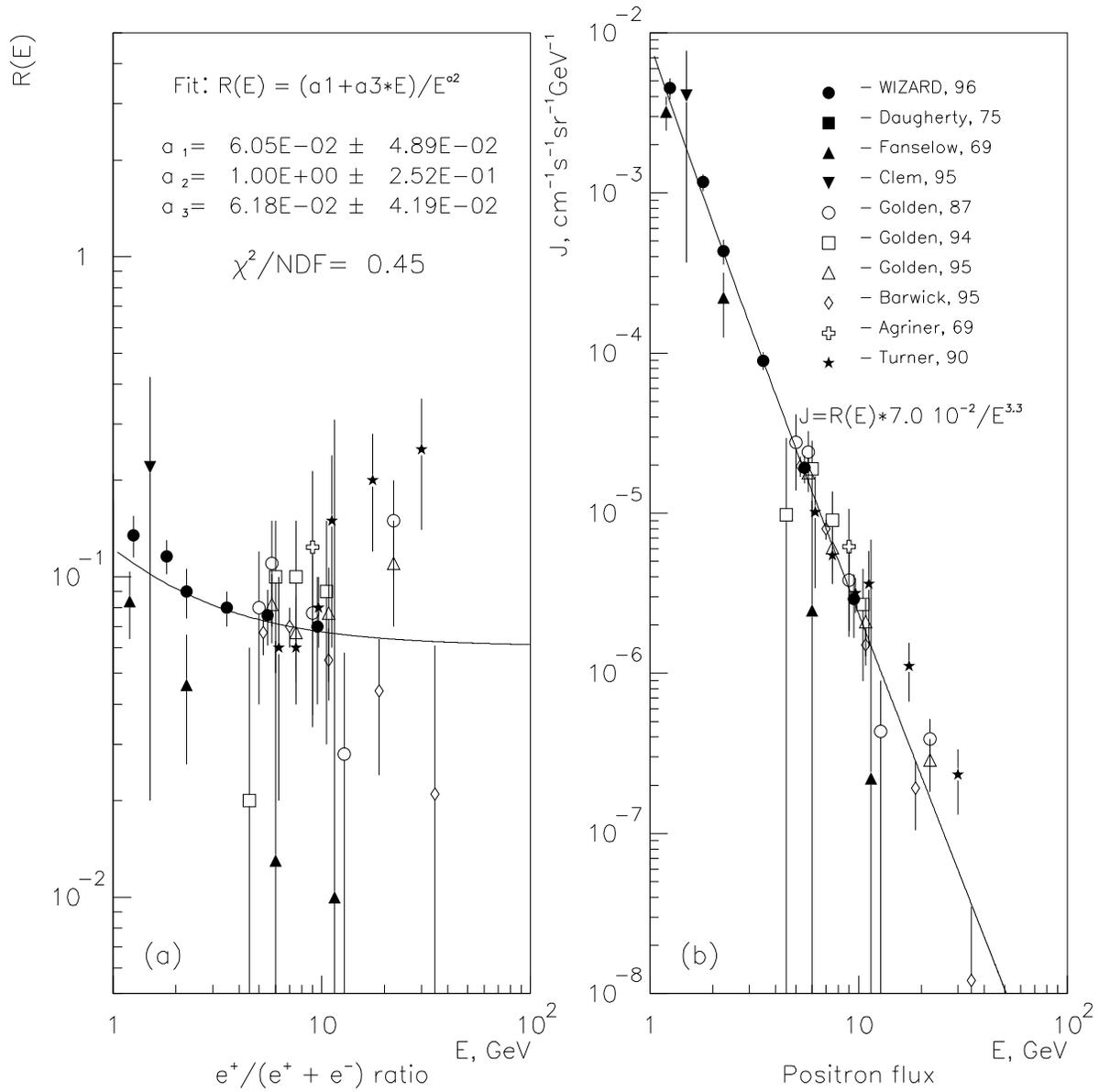,bbllx=0.5cm,bblly=5.0cm,%
bburx=19.0cm,bbury=23.0cm,clip=t,height=16.0cm}
}}
\caption{\label{fitpos}
Measured data for the ratio $e^+/(e^+\,+\,e^-)$ (a)
and for the positron flux (b) together with
the curves resulting from the fit discussed in the text.
}
\end{figure}

\npar
As was noted in \cite{{Fargion95},{Wilczek90}} an important constraint
on $\mnu$ could be obtained by investigating the high energy part
of the positron
spectrum resulting from neutrino annihilation in the Galaxy. 
The flux of positrons
from neutrino annihilation will be the same as in the case of electron 
production but the background conditions are significantly better.
Up to now the positron flux has been measured
only at energies $\le\,30\ GeV$. To estimate the expected positron flux
at higher energies we perfomed a 3--parameter fit to experimental data for
the $e^+/(e^+\,+\,e^-)$ ratio in the form

\begin{equation}
\label{fitform}
R(E)\ =\ \frac{(a_1\,+\,a_2\,E)}{E^{a_3}},
\end{equation}

\noindent with three free parameters $a_1,\,a_2,\,a_3$.
The result of our fit is presented in Fig.\ref{fitpos}.
In the energy region where separate measurements of electrons and positrons 
are available (see Fig.\ref{fitpos}) a significant excess 
of electrons has been 
found. At higher energies we can use either an extrapolation of our fit 
or the theoretical prediction. 
If we assume,  for instance, the validity of the model of the dynamical halo, 
then the ratio $e^+/e^-$ is found to be less than $0.02$ at
$E_e\,>\,30\ GeV$ \cite{Mueller87}.
In this case we expect (see Fig.\ref{mnu100}(b)) a clear signature 
for heavy neutrino annihilation by an edge in the positron spectrum, 
which corresponds to the annihilation process 
\hbox{$\nu\,\bar{\nu}\,\rightarrow\,e^+e^-$.} The detection
of an excess of high energy positrons and a sharp edge of the
positron spectrum (at $E\,=\,\mnu$) would be a clear signature of the 
annihilation of Dirac
neutrinos in the galactic halo \cite{{Fargion95},{Wilczek90}}, since
the direct annihilation of massive Majorana neutrinos into light fermions 
in the Galaxy is severely by suppressed in the nonrelativistic case 
due to angular momentum conservation.

\npar
For the case $\mnu\,=\,300\ GeV$ the resulting electron spectrum is 
well below the background due to a sharp decrease of the neutrino number 
density when the neutrino mass increases above 100 GeV,
$n_{\nu}\,\sim\,\mnu^{-3}$.

\begin{figure}[htb] 
\centerline{\hbox{
\psfig{figure=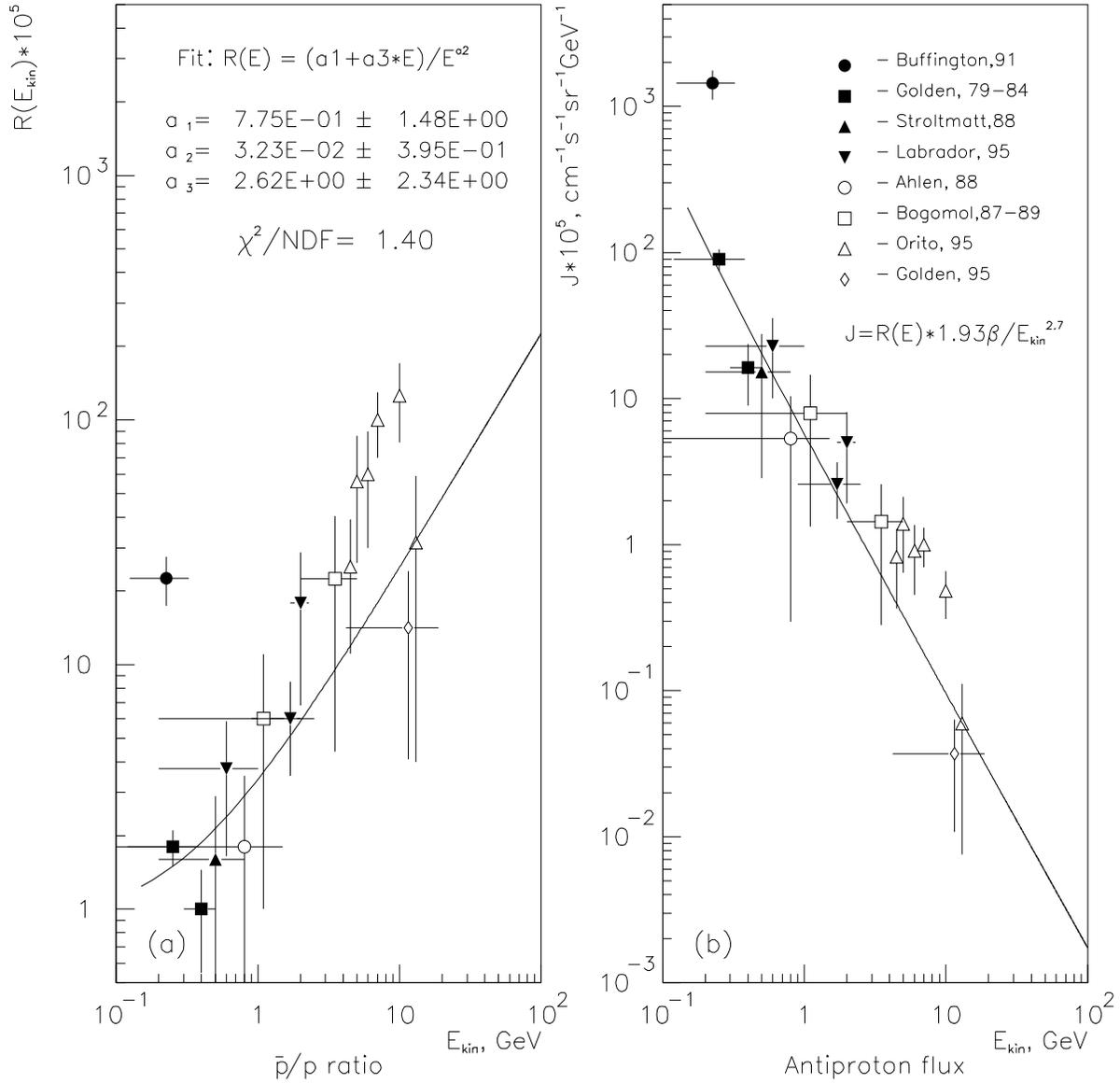,bbllx=0.5cm,bblly=5.0cm,%
bburx=19.0cm,bbury=23.0cm,clip=t,height=16.0cm}
}}
\caption{\label{fitapr}
Measured data for the ratio $\bar p/p$ (a)
and for the antiproton flux (b) together with
the curves resulting from the fit discussed in the text.
}
\end{figure}

\npar
The analysis of the $\bar p$ spectrum is similar to that of electrons and
is based on expression (\ref{elflux}). However, it is important to
note that in the case of antiprotons we can neglect their
energy losses in the galactic disk due to mass suppression of 
synchrotron radiation and inverse Compton scattering
in comparison with the case of electrons. 
Since antiprotons are trapped in the galactic magnetic field and have 
a rather large confinement time of the order $10^7$ yr we can use the same 
expression (\ref{elflux}) for an antiproton flux on the
Earth which appears after a heavy neutrino annihilation and a sequential 
fragmentation of gluons and quarks. At low energies the ratio 
of antiprotons to protons in the cosmic rays is 
extremely low, namely less than about $10^{-4}$ for $E\,\sim\,1\ GeV$ 
(see Fig.\ref{fitapr}). 
Therefore, if a sufficient amount of antiprotons was created 
in the hadronization process the antiproton flux can excess the
background and this could point at $\nu\,\bar{\nu}$ 
annihilation in the halo.

\npar
We present the results of our calculations of antiproton spectra in 
Figs.\ref{mnu100}(c), \ref{mnu300}(c). We ignored a
modulation of the antiproton flux due to solar wind effects 
in the region above $1$ GeV. 
We took the interstellar cosmic ray proton spectrum to be \cite{Ryan72}

\begin{equation}
\label{expprot}
J_p^{exp}\ =\ 1.93\,\left (\frac{v_p}{c}\right )\,E_{kin}^{-2.7}
\ \ \ cm^{-2}\,sr^{-1}\,s^{-1}\,GeV^{-1}.
\end{equation}

\noindent where, $E_{kin}\,=\,E\,-\,M_p$ is a kinetic energy of the proton.

\npar
The dotted line shows our fit to the experimental data for the
$\bar p/p$ ratio (see Fig.\ref{fitapr}) (in the form as Eq.(\ref{fitform}))
multiplied by the proton flux (\ref{expprot}) for the
region above $13$ GeV. As it is seen, the antiproton signal is too small 
to be detected in the case of $\nu\,\bar{\nu}$ annihilation.
Even if we compare our results with standard leaky box model 
prediction for the $\bar p p$ ratio $R_{LB}\,\approx\,10^{-4}$ in the region 
$10\,-\,100\ GeV$ the results are not favourable and cannot impose 
any significant constraint on the model.


\npar
Finally, we discuss the photon spectra resulting from 
$\nu\,\bar{\nu}$ annihilation. 
In the photon case there are no energy losses on the way to the Earth
and the confinement time $\tau$ in eq. (\ref{elflux}) is replaced by an
integral over the line--of--sight which depends on the galactic latitude $b$
and longitude $l$ \cite{{Gunn78},{Ellis88}}:

\begin{equation}
\label{gmflux}
J_{\gamma}^{exp}\ =\ \frac{\vsig}{4\,\pi}\,n_{\nu}^2\,
\frac{dn_{\gamma}}{dE}\,a\,I(b,l),
\end{equation}

\noindent where for the diffuse $\gamma$ component
factor $I(b,l)\,\approx\,1.24$ (high galactic lattitudes) and
$a\,\approx\,10\ kpc$.
The diffuse photon background is not known for $E_{\gamma}\,>\,1\ GeV$. 
Using data from the MeV region and extrapolating to higher energies, 
the photon spectrum is inferred to be \cite{Diehl95}

\begin{equation}
\label{expgam}
J_{\gamma}^{exp}\ =\ 8\cdot10^{-7}\,E_{\gamma}^{-2.7}
\ \ \ cm^{-2}\,sr^{-1}\,s^{-1}\,GeV^{-1}.
\end{equation}

\npar
The results of our $\gamma$ calculations are shown in Fig.\ref{mnu100}(d) 
and Fig.\ref{mnu300}(d). As we see in the case of $\mnu\,=\,100\ GeV$ the
neutrino annihilation leads to an indirect production of photons through 
particle decays above the expected background. However it seems 
that the continuous photon spectrum is not sufficient to give us 
a clear signature of the neutrino annihilation in the galactic halo. 
The difference between the predicted spectrum and the expected background 
is not very significant and the flux is small.

\npar
We should note that our calculations contain significant uncertainties 
in astrophysical assumptions especially about the laws of propagation 
for $\bar p$ and $e^+$ in the Galaxy and the local halo 
density.

\section{Conclusion}

\npar
In this paper we have carried out a detailed analysis of fluxes of
cosmic ray
antiprotons, positrons, electrons and photons to be expected from 
the annihilation of relic heavy stable neutrinos in the galactic halo. 
We used the standard electroweak model including one additional
fermion family 
and the idea of neutrino condensation in the gravitational field of
collapsing matter at the stage of Galaxy formation. Our conclusion is
that it is difficult
on the basis existing experimental data on cosmic ray spectra to constrain 
parameters of heavy neutrinos. However, it seems that future experimental 
study of cosmic rays and especially their $e^+$ and $\gamma$
components in the high 
energy region, could give important information on dark matter particles. 
In particular the detection  of the anomalous positron output corresponding
to the monochromatic annihilation line at the edge of the positron spectrum 
would be a clear 
indication for the annihilation of Dirac particles of dark matter 
in the galactic halo. 
Also the investigation of the continuum spectrum of diffuse 
$\gamma$ rays above $1\ GeV$ could be used in order to constrain 
the parameters of dark matter particles or to provide a confirmation
of a signal seen in other experiments. 
As it was mention in \cite{Ellis88} a possible strategy for future 
experiments is to combine positron detection with those for antiprotons 
and photons.

\npar
Note that direct detection experiments \cite{Ahlen87}
using semiconductor ionization detectors excluded heavy Dirac 
neutrinos as a major component of our galactic halo in the mass 
region $10\,-\, 1000\ GeV$. However, the investigation of cosmic ray
spectra could play an important role in the case of multicomponent 
dark matter and may be used for species (in particular, neutrinos) that give
a negligible contribution to the total energy density of the Universe.

\npar
We also emphasize the possibility of searching for very heavy neutrinos 
in accelerator experiments by the process 
\hbox{$e^+e^-\,\rightarrow\,\nu\,\bar{\nu}\,\,\gamma$}
\cite{Fargion96} which allows investigation in the region $\mnu\,\sim\,m_Z/2$, 
where it is extremely difficult to get cosmological constraints 
on neutrino masses due to a very low neutrino number density in this 
mass region. 
If heavy neutrinos exist there could be also an interesting 
hadronless signature for Higgs meson strahlung production at accelerators 
\hbox{$e^+e^-\,\rightarrow\,Z^0\,H^0\,\rightarrow\,Z^0\,\nu\bar{\nu}
\,\rightarrow\,l^+l^-\,\nu\bar{\nu}$} and this mode could be the dominant one.

\bigskip
\vbox{
\centerline{\bf{Acknowledgement}}
\npar
It is a pleasure to thank Prof. G.Wolf for careful reading of
the manuscript and very useful remarks and comments
and Prof. M.Yu.Khlopov for valuable conversations.

\npar One of the authors (Yu.G.) wants to express his gratitude to DESY
for support during the preparation of this work for publishing.
}

\bigskip
\vbox{
\centerline{Table.1. Total cross sections.}

\bigskip
\centerline{
\begin{tabular}{|l|r|r|}
\hline
& & \\
& $m_{\nu}\,=\,100\ GeV$ & $m_{\nu}\,=\,300\ GeV$ \\
&   $T_f\,=\,3.90\ GeV$  & $T_f\,=\,10.6\ GeV$ \\
& & \\
\hline
Final state & $\sigma_{tot}, (nb)$ & $\sigma_{tot}, (nb)$ \\
\hline
& & \\
$d\bar d$                     & 3.30 & 0.245 \\
$u\bar u$                     & 2.50 & 0.189  \\
$s\bar s$                     & 3.16 & 0.245  \\
$c\bar c$                     & 2.52 & 0.192  \\
$b\bar b$                     & 3.26 & 0.287  \\
$t\bar t$                     & --- &  40.0 \\
$e^-e^+$                      & 0.737 & 0.0554 \\
$\nu_e\bar{\nu}_e$            & 1.48 &  0.110 \\
$\mu^-\mu^+$                  & 0.743 & 0.0554 \\
$\nu_{\mu}\bar{\nu}_{\mu}$    & 1.48  & 0.110 \\
$\tau^-\tau^+$                & 0.724 & 0.0571 \\
$\nu_{\tau}\bar{\nu}_{\tau}$  & 1.48  & 0.110 \\
$W^+W^-$                      & 16.2  & 121 \\
$\Sigma $ over all channels   & 36.2 & 162 \\
& & \\
\hline
\end{tabular}
}
}


\newpage
\centerline{\Large{\bf Appendix.}}

\newcommand{\aem}{\alpha_{em}}
\newcommand{\amnu}{m_{\nu}}
\newcommand{\mnusq}{m_{\nu}^2}
\newcommand{\amf}{m_f}
\newcommand{\mfsq}{m_f^2}
\newcommand{\amz}{m_Z}
\newcommand{\mzsq}{m_Z^2}
\newcommand{\amh}{m_H}
\newcommand{\mhsq}{m_H^2}
\newcommand{\amL}{m_L}
\newcommand{\mLsq}{m_L^2}
\newcommand{\amw}{m_W}
\newcommand{\mwsq}{m_W^2}
\newcommand{\xw}{\sin^2\theta_W}
\newcommand{\WIDZ}{\Gamma_Z}
\newcommand{\WIDH}{\Gamma_H}
\newcommand{\btnu}{\beta_{\nu}}
\newcommand{\Nc}{N_c}
\newcommand{\EI}{e_f}
\newcommand{\AI}{g_A}
\newcommand{\VI}{g_V}
\newcommand{\pk}{(kp)}
\newcommand{\knup}{(k_1p)}
\newcommand{\kk}{(kk_1)}
\newcommand{\pp}{(pp_1)}

\bigskip
\leftline{Differential cross section} 

\begin{eqnarray*}
\label{difcrs}
\frac{d\sigma}{d\,\cos\theta}\ =\ \frac{\beta_f}{32\pi s\btnu}\,
\overline{\vert M\vert^2} \\
\end{eqnarray*}

\bigskip
\leftline{Definitions}

\begin{eqnarray*}
\VI & = &\AI -4\EI\xw \\
G_F & = & \pi\aem/(\sqrt{2}\mwsq\xw) \\
D_Z & = & 1/\left [ (s-\mzsq)^2+\mzsq\WIDZ^2\right ] \\
D_H & = & 1/\left [ (s-\mhsq)^2+\mhsq\WIDH^2\right ] \\
D_L & = & 1/\left [ \mnusq+\mwsq-\mLsq-2\pk\right ] \\
\btnu & = & \sqrt{1-4\amnu^2/s} \\
\beta_f & = & \sqrt{1-4\amf^2/s} \\
\amf & - & \mbox{mass of final particle} \\
\Nc & - & \mbox{number of colours in final state} \\
\end{eqnarray*}

\bigskip
\centerline{\bf Matrix elements squared}

\bigskip
\leftline{1. Reaction $\nu\,\bar {\nu}\,\rightarrow\,f\,\bar f$}

\begin{eqnarray*}
\overline{\vert M_{ff}\vert^2} & = & \Nc\,G_F^2\,\left (F_{ZZ}\,+\,F_{HH}\,+
\,F_{ZH}\right ) \\
F_{ZZ} & = & 2\amw^4/\cos^4\theta_W\,\,D_Z\,
     \biggl [ (\VI +\AI )^2\knup^2+(\VI -\AI )^2\pk^2\ + \\
      &  & \mfsq (\VI^2-\AI^2)\left (\frac{s}{2}-\mnusq\right )\ +
    \ \mnusq\mfsq\AI^2\frac{s}{\mzsq}
     \left (\frac{s}{\mzsq} -2\right )\biggr ] \\
F_{HH} & = & 2\mnusq\,\mfsq\, (s\btnu\beta_f)^2\,D_H \\
F_{ZH} & = & 8\frac{\mwsq}{\cos^2\theta_W}\mnusq\mfsq\VI\,
      \bigl [\pk -\knup \bigr ]\,D_Z\,D_H\,\times \\
    &   & \left [(\mzsq -s- \frac{\WIDZ^2}{4})(\mhsq -s -\frac{\WIDH^2}{4})
          + \amh\amz\WIDH\WIDZ\right ]  \\
\end{eqnarray*}



\bigskip
\leftline{3. Reaction $\nu\,\bar {\nu}\,\rightarrow\,W^+\,W^-$}

\begin{eqnarray*}
\overline{\vert M_{WW}\vert^2} & = & \,G_F^2\amw^4\,\left (F_{ZZ}\,
+\,F_{LL}\,+\,F_{HH}\,+\,F_{ZL}\,+\,F_{ZH}\,+\,F_{HL}\right ) \\
F_{ZZ} & = & D_Z\,\biggl \{ 8\bigl [-(4\pk\knup +7\pk^2 +7\knup^2)\ +
      \ \kk (-\frac{s}{2}-7\mwsq )\bigr ]\ - \\
    &   & \frac{4}{\mwsq} \bigl [2(3\mwsq -s )(2\pk\knup -\mwsq\kk )\ +
      \ 2(4\mwsq -3s )(\knup^2+\pk^2\ - \\
    &   & \kk \pp) - 4(s -\mwsq)^2\kk \bigr ]\ + \\
    & &  \ 2\left (\frac{s}{\mwsq} \right )^2\bigl [-(\pk -\knup )^2 + 
          \kk (\pp -\mwsq )\bigr ]\biggr \} \\
F_{LL} & = & \,D_L^2\,\biggl \{ 16[-2(\mnusq -\pk )\knup
        +\kk (\mnusq -\mwsq )]\ + \\
    &   & \frac{16}{\mwsq} \{2\pk [(\knup -\pp )(\mnusq- \pk )\ + \\
    &   & (\kk -\knup )(\pk -\mwsq )-\knup (\mnusq +\mwsq -2\pk )]\ - \\
    &   & \mwsq [2(\kk -\knup )(\mnusq -\pk )\ -
       \ \kk (\mnusq +\mwsq -2\pk )]\}\ + \\
    &   & \frac{4}{\amw^4} \{2\pk [2(\pk -\mwsq )(2\pk\knup -\mnusq\pp )
          +\mwsq\knup (\mwsq -\mnusq )]\ - \\
    &   & \mwsq [2(\pk -\mwsq )(2\kk\pk -\mnusq\knup )+\kk\mwsq 
          (\mwsq -\mnusq)]\}\biggr \} \\
F_{ZL} & = & 16\,D_Z\,D_L (\mzsq -s -\frac{\WIDZ^2}{4})\,
    \biggl \{\frac{\mnusq s}{2\mzsq} (\knup -\pk -\frac{s}{2}+2\,\mwsq )\ + \\
    &   & \kk (-2\pp +2\mwsq +4\knup +2\pk )\ - \\
    &   & (\mnusq-\pk)(4\pk+2\knup)-4\pk\knup-2\knup^2\ - \\
    &   & \frac{1}{\mwsq} \{(\knup -\pp )[-\pk^2-\knup^2+2\pk\knup\ - \\
    &   & \kk (\mwsq -\pp )]-\pk [2\knup (\pk +2\knup )\ - \\
    &   & \knup (\mwsq +2\pp )+\pk (2\mwsq +\pp ) \\
    &   & -\pp (2\knup +\pk ) - \mnusq (2\mwsq +\pp )]\ +  \\
    &   & (\pk -\mwsq )[\pk^2+\knup^2-2\pk\knup -\kk (\pp -\mwsq )]\ + \\
    &   & \pk [\knup (\pk +\frac{s}{2})+\pk (\knup +\frac{s}{2})\ - \\
    &   & \kk (2\mwsq +\pp )-\mwsq (\frac{s}{2}+\knup )]\ + \\
    &   & \frac{\mwsq}{2} [\kk (4\knup +2\pk -3\pp +3\mwsq )\ + \\
    &   & (\mnusq -\pk )(-5\pk -\knup )-\knup^2-5\pk\knup ]\ + \\
    &   & (s -\mwsq )[\kk (\pp -\mwsq +2\pk )
            -\frac{s}{2}(\mnusq -\pk )\ - \\
    &   & \frac{s}{2}\knup ]-\frac{s\mnusq\mwsq}{2\mzsq} 
           (\pk -\knup -\mwsq +\pp )\}\ + \\
    &   & \frac{1}{2\amw^4} \bigl \{\{2\pk [\pk^2+\knup^2-2\pk\knup
          - \mwsq (\pk -\knup ) \\
    &   & -\pp (\kk -\mnusq )]\ + \\
    &   & \mwsq [\kk (2\pk +\pp -\mwsq )
        - \frac{s}{2}(\mnusq -\pk )-\frac{s}{2}\knup ]\}(s -\mwsq )\ + \\
    &   & \pp \{2\pk [\kk\,(\pp -\mwsq )-\pk^2-\knup^2+2\pk\knup]\ - \\
    &   & \mnusq [(\pk -\knup )\pp +\frac{s}{2}(\pp -\mwsq )]\ - \\
    &   & \mwsq [\kk (\pp -\mwsq )-\pk^2-\knup^2+2\pk\knup ]\}\bigr \}
         \biggr \} \\
F_{HH} & = & 8\mnusq\,D_H \left (\frac{s}{2}-2\mnusq \right )
  \left [\left (\frac{s}{2\mwsq} \right )^2-\frac{s}{\mwsq} +3\right ] \\
F_{HL} & = & 8\mnusq\,D_L\,D_H (\mhsq -s -\frac{\WIDH^2}{4})\,
      \biggl \{ -2[\frac{s}{2}-2\mnusq -\knup +\pk ]\ - \\
    &   & \frac{2}{\mwsq} [(\pk -\knup )(\knup -\pk -\frac{s}{2}
          + \mwsq )-\mwsq (\frac{s}{2}-2\mnusq )]\ +      \\
    &   & \frac{1}{\mwsq} (\frac{s}{2\mwsq} -1)\bigl [
         (\pk -\knup )(\pk -\knup -\mwsq )
          - \pp (\frac{s}{2}-2\mnusq )\bigr ]\biggr \} \\
F_{ZH} & = & -16\mnusq\,D_H\,D_Z\,
 \left [\left (\frac{s}{2\mwsq} \right )^2-3\right ][\pk -\knup]\,\times \\
    &   & \left [(\mzsq -s- \frac{\WIDZ^2}{4})(\mhsq -s -\frac{\WIDH^2}{4})
          + \amh\amz\WIDH\WIDZ\right ] \\
\end{eqnarray*}


\vfill\eject
\vspace*{2cm}

\vfill\eject
\vspace*{2cm}

\vfill\eject
\vspace*{2cm}

\vfill\eject
\vspace*{2cm}

\end{document}